\def\asca{{\it ASCA\/}}

\def\heao1{{\it HEAO-1\/}}

\def\rosat{{\it ROSAT\/}}

\def\eso{ESO~138--G1}
\def\tololo{Tololo~0109--383}

\def\ltsima{$\; \buildrel < \over \sim \;$}
\def\simlt{\lower.5ex\hbox{\ltsima}}
\def\gtsima{$\; \buildrel > \over \sim \;$}
\def\simgt{\lower.5ex\hbox{\gtsima}}

\documentstyle[psfig]{mn}
 
\title[Compton-thick absorption in \tololo\ and \eso]
{Compton-thick X-ray absorption in the Seyfert galaxies \tololo\ and \eso}

\author[M.J. Collinge \& W.N. Brandt]
{\parbox[]{6.5in}{M.J. Collinge\thanks {NASA-supported undergraduate research associate.} and W.N. 
Brandt}\\
\\
Department of Astronomy \& Astrophysics, The Pennsylvania State University, 525 Davey Lab, 
University Park, PA 16802, USA  \\
(collinge@astro.psu.edu and niel@astro.psu.edu) \\
}
 
\begin{document} 

\maketitle


\begin{abstract}
We present analyses of the \asca\ X-ray spectra of two Seyfert galaxies, \tololo\ and \eso. In 
both cases, spectral fitting reveals two statistically acceptable continuum models: Compton 
reflection and 
partial covering. Both spectra have strong iron~K$\alpha$ lines, with equivalent widths greater than 
1.5~keV. These large equivalent widths are suggestive of heavier obscuration than that directly 
indicated by the partial-covering models ($\approx 2\times 10^{23}$~cm$^{-2}$), with the actual 
column densities being `Compton-thick' (i.e. $N_{\rm H}\ga 1.5 \times 10^{24}$~cm$^{-2}$). 
We use the hard X-ray/[O~$\sc iii$] flux 
correlation for Seyferts and data from the literature to provide additional support for this 
hypothesis. Since \tololo\ is known to have optical type~1 characteristics such as broad Balmer 
line components and Fe~$\sc ii$ emission, this result marks it as a notable object.

\end{abstract}

\begin{keywords}
galaxies:~individual:~\tololo~-- galaxies:~individual:~\eso~-- galaxies:~nuclei~-- 
galaxies:~Seyfert~-- X-rays:~galaxies.
\end{keywords}


\section{Introduction}

The issue of absorption in Seyfert nuclei is one that has motivated a great deal of research. For 
example, studying the absorbing gas in Seyfert~2s offers the most straightforward means for 
learning about the putative molecular torus surrounding the central engine, which is a cornerstone 
of the unified model for Seyferts (e.g. Antonucci 1993). These studies are also pertinent to 
research aiming to discover the source of the cosmic X-ray background radiation. In both of these 
cases, a key issue is to understand the distribution of absorbing column densities of the Seyfert 
population. A recent study by Risaliti, Maiolino \& Salvati~(1999) reported that roughly half of 
Seyfert~2s have `Compton-thick' intrinsic absorption columns 
($N_{\rm H}\ga \sigma_{\rm T}^{-1} = 1.5 \times 10^{24}$~cm$^{-2}$), a fraction that is 
higher than indicated by earlier studies that were biased toward bright X-ray sources. Since the 
number of known and well-studied Compton-thick Seyferts is not large, the study of new nearby 
examples is significant. In this paper, we present a study of two Seyferts, \tololo\ (NGC 424) and 
\eso, in which we find evidence for the presence of Compton-thick nuclear absorption. \eso\ is a 
Seyfert~2 (e.g. Alloin et~al. 1992). \tololo\ was originally classified as a Seyfert~2 as well 
(Smith 1975), but later studies (Boisson \& Durret 1986; Durret \& Bergeron 1988; 
Murayama, Taniguchi \& Iwasawa 1998) cast this into doubt. Based on the presence of a broad Balmer 
line component and Fe~$\sc ii$ emission lines in its optical spectrum, Murayama et~al. (1998) 
argued for the type~1 nature of \tololo, settling on a marginal classification between type~1 
and type~2. One of our goals in this study is to use its X-ray properties to gain insight into its 
actual nature.


\setcounter{table}{0}
\begin{table}
\caption{Basic source properties.}
\begin{tabular}{@{}lccccc}
\hline
Object          & $V$         & $z$      & Galactic $N_{\rm H}$\\
\\
\tololo\        & 13.9        & 0.012    & $1.8\times 10^{20}~\rm cm^{-2}~^{\rm a}$\\
\eso\           & 14.3        & 0.0091   & $1.6\times 10^{21}~\rm cm^{-2}~^{\rm b}$\\
\hline
\end{tabular}

$^{\rm a}$Stark et al. (1992) $^{\rm b}$Heiles \& Cleary (1979) \\

\end{table}
%
Both \tololo\ and \eso\ are relatively nearby (see Table~1) and display the so-called coronal 
lines, high-ionization forbidden optical emission lines from species such as [Fe~$\sc vii$], 
[Fe~$\sc x$], and [Fe~$\sc xiv$] (Alloin et~al. 1992; Murayama et~al. 1998). \tololo\ is 
interesting in that it is one of only a few known 
galaxies in which the coronal line region has been observed to be spatially extended. In this 
study, however, we concern ourselves strictly with these galaxies' X-ray 
emission and their optical characteristics (such as [O~$\sc iii$] $\lambda$5007 emission) as they 
relate to their X-ray properties. We adopt $H_0=70$~km~s$^{-1}$~Mpc$^{-1}$ and $q_0={1\over 2}$.
%
%
\begin{figure}
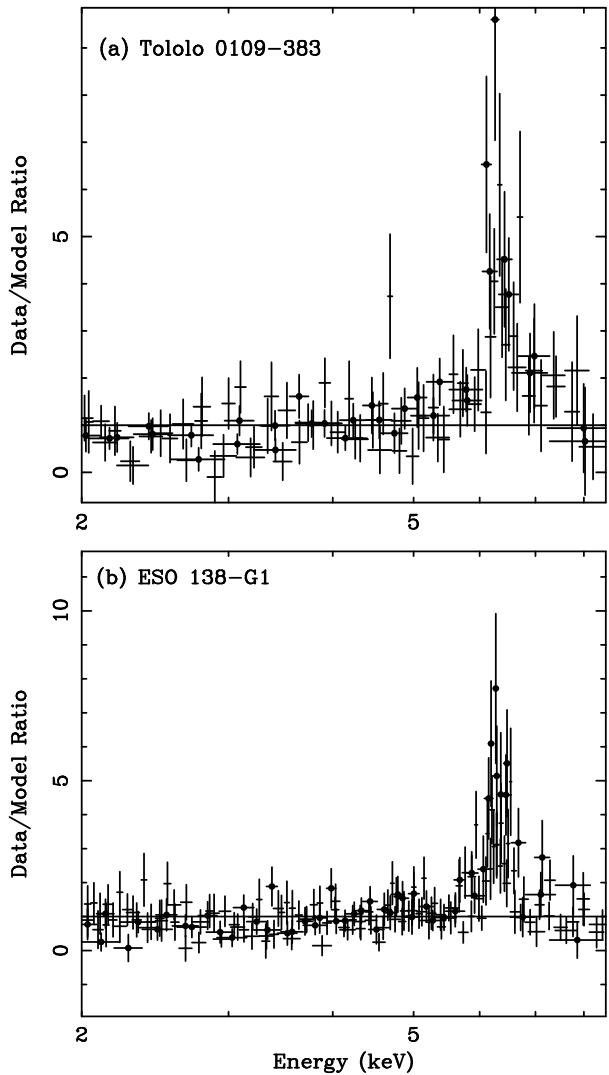

\vspace{0.20 cm} \psfig{figure=tol-line.ps,height=7cm,width=8cm,angle=-90,clip=}
\vspace{0.20 cm} \psfig{figure=eso-line.ps,height=7cm,width=8cm,angle=-90,clip=}
\caption{
Plots of Model~A data-to-model ratio for (a) \tololo\ and (b) \eso. Note the strong residuals due 
to iron~K$\alpha$ emission from 6--7~keV. Dots are SIS data points, and plain crosses are GIS 
data points.
}
\end{figure}
%
%
\section{X-ray observations and analysis}
\subsection{Observations and data reduction}
The \asca\ observations of \tololo\ and \eso\ were performed on 2--3~July~1997 and 
6--7~September~1997, respectively. For our analysis, we used Revision~2 processed data from 
Goddard Space Flight Center, prepared using standard screening criteria (Pier 1997). After 
screening, the respective SIS/GIS exposure times were 34~ks/34~ks for \tololo\ and 27~ks/31~ks 
for \eso. For both objects, we utilized $\sc ximage$ (Giommi, Angelini \& White 1997) to locate 
the sources in the SIS and GIS images. We made use of $\sc xselect$ (Ingham \& Arnaud 1998) to 
reduce the data. 
\subsection{Spectral analysis}
In order to permit $\chi^2$~fitting, we adopted a minimum spectral group size of 15~events per 
data point. For the SIS/GIS detectors, we have included the energy ranges 
0.6--10~keV/0.9--10~keV. We performed simultaneous fitting of the spectra from all four detectors 
for each object, and we constrained all parameters to be the same for each group, aside from the 
absolute model normalizations. We report all equivalent widths and fluxes from the SIS0 detector, 
and all errors at the 90\% confidence level unless otherwise indicated. We have used $\sc xspec$ 
(Arnaud 1996) for all spectral fitting. See Table~2 for a summary of the $\chi^2$~fitting.

We began by fitting the spectra of both objects with a simple power law absorbed by the Galactic 
column density (Model~A). We have allowed for uncertainties of up to 20\% in the Galactic column 
densities; such variations do not materially alter our results. Model~A fails to provide a 
statistically acceptable fit for either spectrum. As shown in Figure~1, there are strong line-like 
residuals above 6~keV in both spectra. In addition, the photon indices are significantly flatter 
than expected for intrinsic Seyfert spectra ($0.99^{+0.16}_{-0.14}$ for \tololo\ and 
$0.55^{+0.12}_{-0.13}$ for \eso). In Model~B, we constrained the photon index of the power law to 
lie in the canonical range for Seyferts; specifically, we required it to be above 1.6 (e.g. Brandt, 
Mathur \& Elvis 1997). We also added a Gaussian emission line to represent iron~K$\alpha$ emission. 
We constrained the line width parameter $\sigma$ to lie below 0.4~keV in order to remain consistent 
with known Seyfert characteristics (e.g. Nandra et~al. 1997). The fit improves somewhat for 
\tololo\ and is comparable to Model~A for \eso\ (see Table~2); neither is statistically 
acceptable. Our next step was to add an intrinsic absorption column (Model~C), but the best-fit 
value of this column density is zero for \tololo, and therefore the fit does not improve. The 
improvement in the fit of the spectrum for \eso\ is slight. 
%
\begin{table*}
\begin{minipage}{150mm}
\caption{Model statistics.}
\begin{tabular}{@{}llcccccc}
\hline
Object Name & Parameter         & Model A              & Model B             & Model C 
         & Model D              & Model E              & Model F\\
\\
\tololo\ & $\chi^2/\sc dof$     & 288 / 171            & 208 / 171           & 208 / 171
         & 164 / 171            & 158 / 171            & 190 / 171\\
 & $1-P(\chi^2\mid\nu)^{\rm a}$ & $5.0\times 10^{-13}$ & $9.5\times 10^{-3}$ & $8.2\times 10^{-3}$
         & 0.44                 & 0.54                 & $5.0\times 10^{-2}$\\
\eso\    & $\chi^{2}/\sc dof$  & 434 / 289            & 434 / 289           & 412 / 289 
         & 288 / 289            & 281 / 289            & 376 / 289\\
         & $1-P(\chi^2\mid\nu)$ & $8.3\times 10^{-30}$ & $1.2\times 10^{-8}$ & $4.4\times 10^{-7}$ 
         & 0.34                 & 0.46                 & $6.8\times 10^{-5}$\\
\hline
\end{tabular}

$^{\rm a}P(\chi^2\mid\nu)$ is the probability that the observed chi-square for a correct model 
should be less than a value $\chi^2$ (Press et~al. 1989).

\end{minipage}
\end{table*}
%
%

We were able to achieve statistically acceptable fits of both spectra using models consisting of 
absorption by the Galactic column densities of Compton-reflected continua (`pexrav') plus iron 
lines (Model~D). This type of model is often found to 
represent well the spectra of absorbed Seyferts. Because of the poor signal-to-noise 
of our data, we constrained the element abundances to be solar, and we fixed the power-law cutoff 
energy to be far outside the \asca\ band at 1000~keV. In Table~3 we list the relevant fit parameters. 
The high values of the reflection scaling factors indicate that reflected components dominate both 
spectra, and the photon indices, although poorly constrained, are consistent with reasonable values. 
We note that the $\approx 6.4$~keV lines are consistent with neutral iron~K$\alpha$ emission and 
that the statistical degradation of the fit is not large if we constrain the line in the spectrum of 
\tololo\ to be narrow ($\sigma = 0.05$~keV).

Statistically acceptable fits can also be achieved using models consisting of power laws plus iron 
lines absorbed by the Galactic column densities and partial-covering (intrinsic) absorption columns 
(Model~E). Like reflection models, partial covering is also often found to be important in the 
spectra of Seyferts with 
obscuration. The photon indices and line equivalent widths obtained from this model are consistent 
with those obtained from Model~D, and again the line energies and widths are consistent with neutral, 
narrow iron K$\alpha$ emission. We shall later discriminate between Models D and E based on physical 
arguments.

As a means of checking the general robustness of these results, we also considered alternate models 
consisting of Raymond-Smith plasma components added to Model~C (Model~F). Physically the plasma 
component represents thermal gas emission such as is often associated with starburst activity 
(e.g. Ptak et~al. 1999). For each spectrum, the fit is a significant improvement over Model~B; 
however, the fits are still not statistically acceptable.
%
%
\begin{figure}
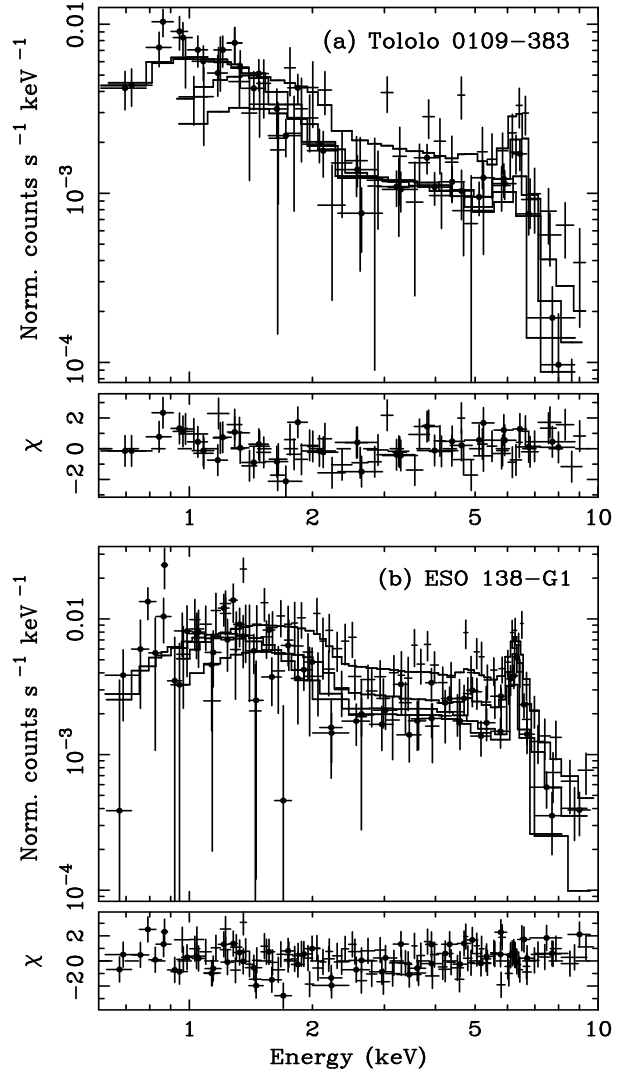

\vspace{0.20 cm} \psfig{figure=tol-mod.ps,height=7cm,width=8cm,angle=-90,clip=}
\vspace{0.20 cm} \psfig{figure=eso-mod.ps,height=7cm,width=8cm,angle=-90,clip=}
\caption{
\asca\ SIS (dots) and GIS (plain crosses) spectra of (a) \tololo\ and (b) \eso. Model~D has been 
fit to the data. The ordinates for the lower panels (labeled $\chi$) show the fit residuals in 
terms of sigmas with error bars of size one.
}
\end{figure}
%
%
\begin{table}
\caption{Best-fit Model~D parameters with 90\% ($\Delta\chi^2=2.71$) errors, X-ray fluxes, and 
[O~$\sc iii$] fluxes.}
\begin{tabular}{@{}lcc}
\hline

Parameter$^{\rm a}$                         & \tololo\               & \eso\                  \\
\\
Refl. scaling factor                        & $88^{+120}_{-14}$      & $70^{+130}_{-32}$      \\
Photon index ($\Gamma$)                     & $2.3^{+0.2}_{-0.2}$    & $2.1^{+0.4}_{-0.4}$    \\
Line energy (keV)                           & $6.43^{+0.09}_{-0.10}$ & $6.36^{+0.05}_{-0.05}$ \\
Line width ($\sigma$) (keV)                 & $0.20^{+0.28}_{-0.12}$ & $0.07^{+0.09}_{-0.07}$ \\
Equivalent width (keV)                      & $1.6^{+0.4}_{-0.5}$    & $1.7^{+0.4}_{-0.4}$    \\
$F_{0.6-2}^{\rm b}$                         & $2.0\times 10^{-13}$   & $1.8\times 10^{-13}$   \\
$F_{2-10}^{\rm b}$                          & $1.2\times 10^{-12}$   & $1.8\times 10^{-12}$   \\
$F_{[\rm O\ \sc iii]}^{\rm c}$              & $2.3\times 10^{-13}$   & $9.3\times 10^{-13}$   \\

\hline
\end{tabular}

$^{\rm a}$All fluxes are reported in erg~cm$^{-2}$~s$^{-1}$. $^{\rm b}$SIS0 observed 0.6--2~keV 
and 2--10~keV fluxes. $^{\rm c}[\rm O\ \sc iii]\ \lambda 5007$ emission, corrected for 
Galactic extinction. From Murayama et~al. (1998) and Schmitt \& Storchi-Bergmann 
(1995), respectively. The value reported for \eso\ is the total flux within $5^{\prime\prime}$ of 
the nucleus.

\end{table}
%
%
\begin{figure}
\vspace{0.20 cm} \psfig{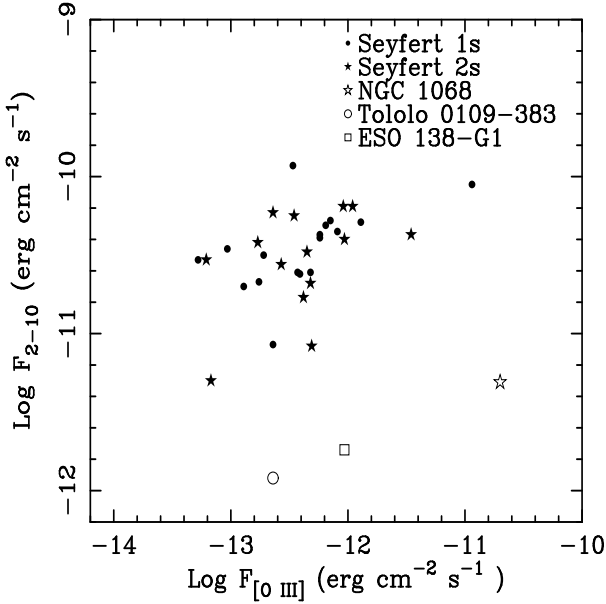} 
\caption{
$F_{2-10}$ versus $F_{[\rm O\ \sc iii]}$ for the Seyferts in the Mulchaey et~al. (1994) sample 
plus \tololo\ and \eso. We have updated the 2--10 keV flux for NGC~4051 (a Seyfert~1) using the 
average flux reported by Guainazzi et~al. (1996). The object in the lower right-hand corner, 
NGC~1068, is a well-known Compton-thick Seyfert~2 (e.g. Matt et~al. 1997).
}
\end{figure}
%
%
\section{Discussion and Conclusions}
\subsection{The iron K$\balpha$ lines and variability}
Iron K$\alpha$ lines are produced through the reprocessing of primary X-rays and become strongest 
in equivalent width when the primary X-ray continuum is suppressed in the neighborhood of 
6.4--6.97~keV. The large iron K$\alpha$ equivalent widths (greater than 1.5~keV) and 
flat apparent spectral continua (see Section 2.2) that we observe in \tololo\ and \eso\ 
indicate that reprocessing is important in these sources (e.g. Matt, Brandt \& Fabian 
1996), a deduction which seems to be confirmed by the good fits of the spectra that we achieved 
using Compton-reflection models (Model~D). The large reflection scaling factors in those fits 
indicate that the reflected components dominate any direct components of the nuclear emission to a 
great extent. We achieved equally good fits, however, using partial-covering models. We must 
therefore discriminate between the two models based on physical arguments.

Matt et~al. (1996) describe two reprocessing mechanisms that can be significant in 
Seyferts, both of which can produce very strong iron~K$\alpha$ emission. In the first case, the 
iron emission occurs at the inner surface of the circumnuclear absorbing material (the putative 
torus). The viewing angle can be such that the central source is obscured but the inner wall of 
the torus is not. In this case, predominantly neutral emission is expected and iron~K$\alpha$ 
equivalent widths can be up to a few~keV. Our Model~D is entirely consistent with this 
theoretical mechanism. The second scenario places the iron emission in 
optically thin gas in the nuclear region (the `warm mirror'). As long as the direct X-ray 
emission is obscured, the iron~K$\alpha$ equivalent width can be as high as a few~keV, but in 
this case the line is expected at higher energies (6.7--6.97~keV), representing emission from 
very highly ionized iron. If we interpret the partial-covering column densities as representing 
absorbing gas along some lines of sight through the scattering regions, our Model~E is consistent 
with this scenario, {\it except for the energies of the iron~K$\alpha$ lines}. These energies 
are inconsistent at greater than the 90\% confidence level with emission from the highest 
ionization states of iron. Thus we can conclude that a scattered component is not highly 
significant in the spectra of either \tololo\ or \eso.

Based on these considerations, it is likely that we are primarily seeing reprocessing by the 
inner surface of a circumnuclear absorber (torus). In order to get the large equivalent widths 
we observe (greater than 1.5~keV), however, an absorption column density is 
required that is significantly larger than the intrinsic column densities indicated by Model~E. 
A column density of $2\times 10^{23}$~cm$^{-2}$ (of the order we fit 
in \tololo\ and \eso) absorbs only $\approx 25$\% of the X-ray flux at 6.4~keV. To get 
such a large equivalent width in the 6--7~keV range, the 
intrinsic absorption along a direct line of sight to the X-ray source must be essentially 
Compton-thick (e.g. Matt et~al. 1996).

We have also analyzed \rosat\ data for \tololo\ and \eso\ in order to check for flux variability 
between the \rosat\ and \asca\ observations of these objects. \tololo\ was observed by the \rosat\ 
PSPC on 2--3~July~1992 (Rush \& Malkan 1996), and \eso\ appears in the \rosat\ HRI field of SN1990W, 
which was observed on 11--12~September~1995. Hence the variability timescales we are probing are 
roughly 5~years and 2~years, respectively. Significant low-energy variability would 
place a constraint upon the size of the scattering/reflecting regions. Based on our analysis 
of the data and the $F_{0.1-2}$ for \tololo\ quoted by Rush \& Malkan (1996), however, we find 
no evidence for 
significant variability between the \rosat\ and \asca\ observations of either source. The fluxes 
are consistent to within the cross-calibration uncertainties of the detectors (e.g. Iwasawa, 
Fabian \& Nandra 1999).
%
%
\subsection{Additional evidence for Compton-thick absorption}
We now consider additional evidence supporting the conclusion that the primary nuclear sources in 
\tololo\ and \eso\ are completely obscured in the \asca\ band. Mulchaey et~al. (1994) reported 
correlations between $F_{[\rm O\ \sc iii]}$ (see Table~3) and $F_{2-10}$ for Seyferts. They found 
the following relationships (the reported error ranges are one standard deviation):
\\
\\
log$(F_{[\rm O\ \sc iii]}/F_{2-10})=-1.89\pm 0.50$~~~(Seyfert~1s)\\
log$(F_{[\rm O\ \sc iii]}/F_{2-10})=-1.76\pm 0.62$~~~(Seyfert~2s)\\
\\
A simple calculation using the fluxes from Table~3 yields the values 
log$(F_{[\rm O\ \sc iii]}/F_{2-10})=-0.72$ for \tololo\ and 
log$(F_{[\rm O\ \sc iii]}/F_{2-10})=-0.29$ for \eso. These numbers fall well outside the $1\sigma$ 
and $2\sigma$ ranges, respectively, of even the Seyfert~2 relationship (see Figure~3). 
This provides further support for the obscuration of most of the direct X-rays from these 
sources below 10~keV, strengthening the conclusion that the intrinsic absorption in these galaxies 
is Compton-thick. To emphasize this point, we have 
used the Seyfert~2 relationship to predict $F_{2-10}$ for the two galaxies from 
$F_{[\rm O\ \sc iii]}$ (e.g. Turner et~al. 1997), and we compare these to the observed values 
of $F_{2-10}$. Using the 
Table~3 fluxes, we obtain values of $11^{+35}_{-8}$ for \tololo\ and $30^{+94}_{-23}$ 
for \eso\ for the ratios of predicted to observed fluxes. We note that this is a somewhat 
conservative estimate for \tololo, to which we might also justifiably apply the Seyfert~1 
relationship.
%
%
\subsection{\tololo, a Compton-thick type~1 Seyfert}
Although the optical properties of \tololo\ support a Seyfert~1 or intermediate nature 
(see 
Section~1), its inferred absorption column density and iron K$\alpha$ line are more like those of 
Seyfert~2s. It seems clear that in the 
case of this object, the simplified picture of a type~1 or type~2 nature falls somewhat short of 
the truth. Nevertheless, we argue that \tololo\ is a Compton-thick Seyfert; this seems to be the 
only plausible explanation for its observed X-ray properties. The combination of its type~1 
optical properties with this heavy intrinsic absorption marks it as a member of a peculiar 
class of objects. It may well serve as a nearby archetype for the heavily obscured type~1 objects 
recently found in deep X-ray surveys (e.g. Comastri et~al. 2000; also see Brandt, Laor \& Wills 2000).
As we have only placed a lower limit on the intrinsic absorption column in 
this object, we point out that the remaining uncertainty in its actual value 
should be resolvable by an observation further into the hard X-ray band by a 
satellite such as $\it BeppoSAX$. Finally, we comment that the existence of type~1 objects with 
heavy X-ray obscuration, such as \tololo\ and Broad Absorption Line QSOs (e.g. Gallagher et~al. 
1999), should give pause to those who would classify an object as type~2 based solely on a hard 
X-ray spectrum.
%
%
\section*{Acknowledgments}
We gratefully acknowledge financial support from NASA grant NAG5-7256 and the Barry M. Goldwater 
Scholarship and Excellence in Education Foundation (MJC) and NASA LTSA grant NAG5-8107 (WNB). We 
thank the referee, Giorgio Matt, for constructive comments. This
research has made use of data obtained through the High Energy Astrophysics Science Archive
Research Center Online Service, provided by NASA's Goddard Space Flight Center.

After this paper was submitted, we learned that Matt et~al. (2000) have confirmed the Compton-thick 
nature of the absorber in \tololo\ using data from {\it BeppoSAX}.
%
%
{}


\bsp

\end{document}